\documentclass[aps,twocolumn,showpacs]{revtex4}

\usepackage{graphics,epsfig,graphicx}

\begin{document}

\title{Multiple filamentation induced by input-beam ellipticity}

\author{Audrius Dubietis and Gintaras Tamo\u{s}auskas}

\affiliation{Vilnius University, Department of Quantum
Electronics, Saul\.{e}tekio Ave. 9, bldg. 3, LT-2040 Vilnius,
Lithuania}

\author{Gadi Fibich}

\affiliation{Department of Applied Mathematics, Tel Aviv
University, Tel Aviv 69978, Israel}

\author{Boaz Ilan}

\affiliation{Department of Applied Mathematics, University of
Colorado, Campus Box 526, Boulder, Colorado 80309-0526}

\begin{abstract}
The standard explanation for multiple filamentation (MF) of
intense laser beams has been that it is initiated by input beam
noise (modulational instability). In this study we provide the
first experimental evidence that MF can also be induced by input
beam ellipticity. Unlike noise-induced beam breakup, the MF
pattern induced by ellipticity is reproducible shot to shot.
Moreover, our experiments show that ellipticity can dominate the
effect of noise, thus providing the first experimental methodology
for controlling the MF pattern of noisy beams. The results are
explained using a theoretical model and simulations.
\end{abstract}


\maketitle

The propagation of high-power ultrashort pulses through the
atmosphere is currently one of the most active areas of research
in nonlinear optics, with potential applications such as remote
sensing of the atmosphere and lightning
control~\cite{applications}. In experiments, narrow filaments of
typical width of 100$\mu m$ have been observed to propagate over
distances of hundreds of meters, i.e., over many Rayleigh lengths.
The stability of a single filament over such long distances is
nowadays known to be the result of the dynamic balance between the
focusing Kerr nonlinearity, diffraction and the defocusing effect
of plasma formation due to multiphoton ionization. The initial
stage of propagation during which filaments are formed, however,
is much less understood. In particular, since in these experiments
the laser power is many times the critical power for
self-focusing, a single input beam typically breaks-up into
several long and narrow filaments, a phenomenon known as multiple
filamentation (MF). Since MF involves a complete breakup of the
beam cylindrical symmetry, it has to be initiated by a
symmetry-breaking mechanism. The standard explanation for MF in
the Literature has been that it is initiated by input beam
noise~\cite{Bespalov-66}, see also Ref.~\cite{Berge} for a review.
Since noise is, by definition, random, this implied that the MF
pattern would be different from shot to shot, i.e., the number and
location of the filaments is unpredictable. This constitutes a
serious drawback in applications where precise localization is
crucial (e.g., laser eye surgery) or in experiments where one
wants to measure the filament properties (power, transverse
profile, etc.) after some propagation distance. Unfortunately,
noise is always present in such high-power lasers, and is not easy
to eliminate to a degree that will lead to a deterministic MF
pattern.

Recently it was predicted theoretically that input beam
ellipticity can also lead to MF~\cite{circular}. In this case the
MF pattern is deterministic, i.e., reproducible from shot to shot.
{\em In this study we provide the first experimental evidence
that input beam ellipticity can indeed induce a deterministic MF
pattern}. Moreover, although a certain amount of noise is present
in our beam, we observe that the MF pattern is nearly identical
from shot to shot. This shows that sufficiently large ellipticity
can dominate noise in the determination of the MF pattern. In
other words, rather then trying to eliminate noise, one can
control the MF pattern by adding sufficiently large ellipticity to
a noisy input beam.

Most recent experimental studies of MF of intense laser beams have
been performed in connection with atmospheric
propagation~\cite{Mlejnek, Schwarz, Tzortzakis, Chin}. Despite
some important differences (nonlinear response, dispersion, etc.)
between gases and condensed media, it is expected that the
physical processes leading to MF are very similar in both cases.
Indeed, some of the present authors have recently demonstrated
self-guided propagation of femtosecond light pulses in water for
distances exceeding several Rayleigh lengths~\cite{Dubietis-03}.
In the experiments reported in this study, we increase the
incident beam power and modify its spatial parameters, resulting
in MF in water.

A 170-fs, 527-nm pulse was provided by second-harmonic compressed
Nd:glass laser system (TWINKLE, Light Conversion Ltd., Lithuania)
operated at 33 Hz repetition rate. Spatially filtered beam was
focused into $\sim 85 \mu $m FWHM beam waist at the entrance of
water cell by means of f=+500 mm lens. Incident energy was varied
by means of a half-wave plate and a polarizer. The focused beam
has a small intrinsic ellipticity, which was evaluated as a
parameter e=a/b=1.09. Highly elliptical beam (e=2.2) was formed by
inserting slightly off-axis iris into the beam path. The output
face of the water cell was imaged onto the CCD camera (Pulnix
TM-6CN and frame grabber from Spiricon, Inc., Logan, Utah) with
7$\times$ magnification by means of an achromatic objective (f=+50
mm).

\begin{figure}
    \centerline{\scalebox{0.95}{\includegraphics{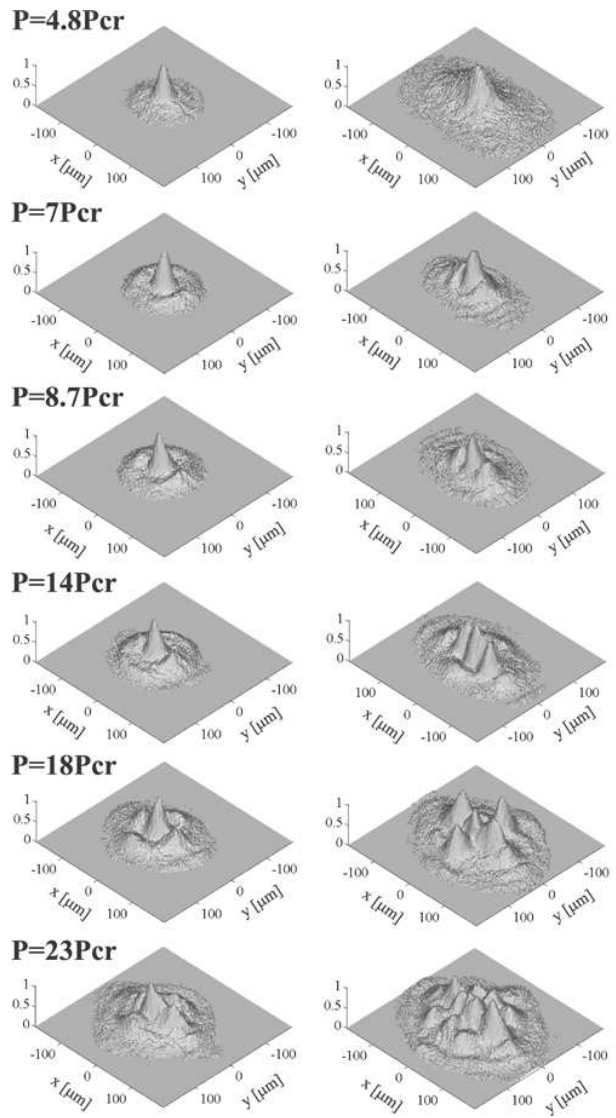}}}
\caption{Normalized 3D views of filamentation patterns at z=31 mm
recorded with circular incident beam (e=1.09, left panel) and
elliptical incident beam (e=2.2, right panel). The major axis of
the ellipse lies along the x-axis of the plots.
$P_{cr}=3.77{\lambda^2}/(8\pi n n_2)=1.15 MW$}
  \label{fig:mf}
\end{figure}

In the first series of experiments we recorded transverse
distribution patterns at fixed propagation length z=31 mm ($\sim
0.7L_{DF}$, $L_{DF}= n k_0 r_0/2$) as we increased the incident
power, see Fig.~\ref{fig:mf}. Two cases were examined; a
near-circular input beam (e=1.09) and an elliptic beam (e=2.2).
Several important conclusions can be drawn: 1) The threshold power
for MF is much less for the elliptic beam, 2) The number of
filaments increases with input power, 3) At power levels
moderately above the threshold for MF, in addition to the central
filament, there are two filaments along the major axis of the
ellipse. At higher powers there are additional filaments in the
perpendicular direction. At even higher powers (P=23$P_{cr}$) one
can observe a quadruple of filaments along the bisectors of the
major and minor axes. 4) MF starts as nucleation of an annular
ring, which contains the power that was not trapped in the central
filament (this is more evident for e=1.09). 5) Since the MF
patterns shown in Fig.~\ref{fig:mf} were reproducible from shot to
shot, they were not induced by random noise. 6) Investigation of
{\em dynamics} of the MF structure (data not presented here)
showed that it is robust in terms of propagation, i.e., after an
initial transient each of the filaments propagates as an
independent entity.

In Fig.~\ref{fig:mf} we observe that the side filaments are always
pairs located symmetrically along the major and/or minor axis,
and/or quadruples located symmetrically along the bisectors of the
major and minor axes. This observation can be explained based on
the following symmetry argument. Consider an elliptic input beam
of the form $E_0(x,y,t) = F(x^2/a^2+y^2/b^2,t)$. Since the medium
is isotropic, the electric field E should be symmetric with
respect to the transformation $x \rightarrow -x$ and  $y
\rightarrow -y$. Therefore, if the filamentation pattern is
induced by input beam ellipticity, it can only consist of a
combination of 1) a single on-axis central filament, 2) pairs of
identical filaments located along the ellipse major axis at $(\pm
x, 0)$, 3) pairs of identical filaments located along the minor
axis at $(0, \pm y)$, and 4) quadruples of identical filaments
located at $(\pm x, \pm y)$.

Whereas ellipticity decreases the threshold power for MF, it
increases the threshold power for the formation of a single
filament. Indeed, the threshold for observing a single filament at
z=31 mm were 6$P_{cr}$ and 4.9$P_{cr}$ for the elliptic and the
near-circular beams, respectively. This $\sim20\%$ increase is in
good agreement with the theoretical prediction for the increase in
the threshold power for collapse (of cw beams) due to beam
ellipticity~\cite{elliptic-00}.

\begin{figure}
    \centerline{\scalebox{0.95}{\includegraphics{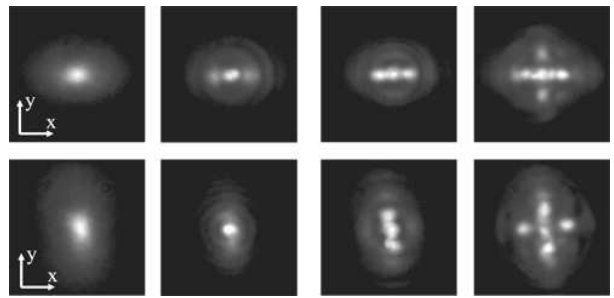}}}
\caption{ CCD camera images of the filamentation patterns of
elliptical beams with different transverse orientation (denoted by
x and y axes) at z=31mm. Image area is $330 \times 330 \mu m^2$,
incident power is 5, 7, 10, and 14$P_{cr}$, as seen from left to
right.} \label{fig:rotation}
\end{figure}

In the experiment shown in Fig.~\ref{fig:rotation} we produced two
input beams with the same ellipticity parameter (e=2.2), but with
different orientations in the transverse plane. In both cases we
observe that the beam is elliptic and still focusing at
P=5$P_{cr}$, a single central filament at P=7$P_{cr}$, an
additional pair of comparable-power secondary filaments along the
major axis of the ellipse at P=10$P_{cr}$, and a second pair of
weaker filaments in the perpendicular direction at P=14$P_{cr}$.
{\em The rotation of the filamentation pattern with the ellipse
rotation thus confirms that the MF in these experiments is indeed
induced by the intrinsic beam ellipticity.}

We recall that it was recently shown that polarization effects
could also lead to reproducible MF pattern \cite{polarization}. In
that case, however, the orientation of the filamentation pattern
is determined by the direction of linear polarization. To check
that, we changed the direction of linear polarization of the
incident beam and verified that it has no effect on the
orientation of the MF pattern. Indeed, polarization effects are
important only when the radius of a single filament becomes
comparable with the wavelength. This is not the case in our
experiments, as the FWHM diameter of a single filament is $\sim 20
\mu m$.

\begin{figure}
    \centerline{\scalebox{1}{\includegraphics{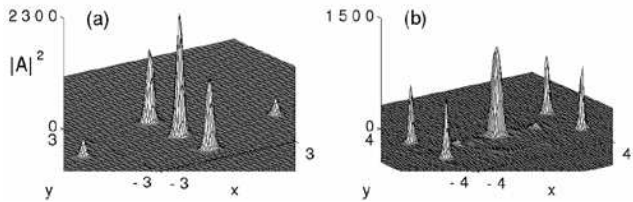}}}
    \caption{Solution of Eq.~(\ref{eq:NLS}) with $\epsilon_{sat} = 0.005$
      and $P=66 P_{cr}$. (a)~ $e=1.09$ and $z=0.9$. (b)~ $e=2.2$ and $z=0.7$.}
        \label{fig:model}
\end{figure}

In our simulations we used a simpler model of propagation of cw beams
in a medium with a saturable nonlinearity, i.e.,
\begin{eqnarray}
  \label{eq:NLS}
  \lefteqn{
    i A_z(z,x,y) + \Delta A+\frac{|A|^2}{1+ \epsilon_{sat} |A|^2} A = 0, 
   \hspace{0.2in}  } \\*[2mm] 
  && \hspace{0.2in} 
  A(0,x,y) = c e^{-x^2/e^2-y^2}. \nonumber
\end{eqnarray}
This model is considerably simpler than the physics governing
propagation of intense ultrashort pulses in water. Nevertheless,
numerical simulations of equation~(\ref{eq:NLS}) reproduced the
same qualitative features observed experimentally. For example, in
Fig.~\ref{fig:model}(a) the MF pattern consists of a strong
central filament, a pair of filaments along the minor axis, and a
second pair of weaker filaments along the major axis. In
Fig.~\ref{fig:model}(b) the MF pattern consists of a central
filament, a quadruple of filaments along the lines $y = \pm
0.37x$, and a pair of very weak filaments along the major axis.
These simulations, therefore, suggest that {\em MF induced by
ellipticity is a generic phenomenon that does not depend on the
specific optical properties of the medium (air, water, silica,
etc.) or on pulse duration}.

In conclusion, we have demonstrated for the first time that input
beam ellipticity can lead to MF. Unlike noise-induced MF, the
filamentation pattern is reproducible and consists only of a
central filament and/or pairs of identical filaments lying along
the major and/or minor axes of the ellipse, and/or quadruples of
identical filaments along the bisectors of the major and minor
axes. The effect of ellipticity on MF seems to be generic,
i.e., independent of the optical properties of the medium.
Since a certain amount of astigmatism is always present in
experimental setups, this observation may explain previous MF
experiments, in which the filamentation pattern was reproducible.
In addition, this study shows that one can overcome the random
nature of noise and control the MF pattern simply by adding large
ellipticity to the input beam.

\bibliographystyle{unsrt}

\end{document}